\documentclass[twocolumn]{aastex63}

\newcommand{\kms}{${\rm km\,s^{-1}}$}

\newcommand{\ccc}{$^{13}$C$^{18}$O}
\newcommand{\ce}{C$^{18}$O}
\newcommand{\cseven}{C$^{17}$O}
\def\farcs{\hbox{$.\!\!^{\prime\prime}$}}
\shorttitle{Excess C/H Gas from Ice Pebble Drift}
\shortauthors{Zhang et al.}
\graphicspath{{./}{figures/}}
\begin{document}

%\title{Super-stellar C/H Gas in the Giant Planet Formation Zone of the HD 163296 Protoplanetary Disk}
\title{Excess C/H in Protoplanetary Disk Gas from Icy Pebble Drift across the CO Snowline}

\correspondingauthor{Ke Zhang}
\email{kezhang@umich.edu}

\author[0000-0002-0661-7517]{Ke Zhang}
\affiliation{Department of Astronomy, University of Michigan, 
323 West Hall, 1085 S. University Avenue, 
Ann Arbor, MI 48109, USA}
\affiliation{Hubble Fellow}

\author[0000-0003-4001-3589]{Arthur D. Bosman}
\affiliation{Department of Astronomy, University of Michigan,
323 West Hall, 1085 S. University Avenue,
Ann Arbor, MI 48109, USA}

\author[0000-0003-4179-6394]{Edwin A. Bergin}
\affiliation{Department of Astronomy, University of Michigan,
323 West Hall, 1085 S. University Avenue,
Ann Arbor, MI 48109, USA}

%% Note that the \and command from previous versions of AASTeX is now
%% depreciated in this version as it is no longer necessary. AASTeX 
%% automatically takes care of all commas and "and"s between authors names.

%% AASTeX 6.3 has the new \collaboration and \nocollaboration commands to
%% provide the collaboration status of a group of authors. These commands 
%% can be used either before or after the list of corresponding authors. The
%% argument for \collaboration is the collaboration identifier. Authors are
%% encouraged to surround collaboration identifiers with ()s. The 
%% \nocollaboration command takes no argument and exists to indicate that
%% the nearby authors are not part of surrounding collaborations.

%% Mark off the abstract in the ``abstract'' environment. 
\begin{abstract}

The atmospheric composition of giant planets carries the information of their formation history.  Superstellar C/H ratios are seen in atmospheres of Jupiter, Saturn, and various giant exoplanets. Also, giant exoplanets show a wide range of C/O ratio. To explain these ratios, one hypothesis is that protoplanets accrete carbon-enriched gas when a large number of icy pebbles drift across the CO snowline. Here we report the first direct evidence of an elevated C/H ratio in disk gas. We use two thermo-chemical codes to model the \ccc, \cseven, and \ce~(2-1) line spectra of the HD 163296 disk. We show that the gas inside the CO snowline ($\sim$70\,au) has a C/H ratio of 1-2 times higher than the stellar value. This ratio exceeds the expected value substantially, as only 25-60\%~of the carbon should be in gas at these radii. Although we cannot rule out the case of a normal C/H ratio inside 70 au, the most probable solution is an elevated C/H ratio of 2-8 times higher than the expectation. Our model also shows that the gas outside 70\,au has a C/H ratio of 0.1$\times$ the stellar value. This picture of enriched C/H gas at the inner region and depleted gas at the outer region is consistent with numerical simulations of icy pebble growth and drift in protoplanetary disks. Our results demonstrate that the large-scale drift of icy pebble can occur in disks and may significantly change the disk gas composition for planet formation. 

% the gas outside the CO snowline ($\sim$70\,au) has a C/H ratio of 0.1$\times$ the ISM value. And are hard to be explained by planetesimal contamination alone. 
\end{abstract}

%% Keywords should appear after the \end{abstract} command. 
%% See the online documentation for the full list of available subject
%% keywords and the rules for their use.
% \keywords{editorials, notices --- 
% miscellaneous --- catalogs --- surveys}

%% From the front matter, we move on to the body of the paper.
%% Sections are demarcated by \section and \subsection, respectively.
%% Observe the use of the LaTeX \label
%% command after the \subsection to give a symbolic KEY to the
%% subsection for cross-referencing in a \ref command.
%% You can use LaTeX's \ref and \label commands to keep track of
%% cross-references to sections, equations, tables, and figures.
%% That way, if you change the order of any elements, LaTeX will
%% automatically renumber them.
%%
%% We recommend that authors also use the natbib \citep
%% and \citet commands to identify citations.  The citations are
%% tied to the reference list via symbolic KEYs. The KEY corresponds
%% to the KEY in the \bibitem in the reference list below. 

\section{Introduction} \label{sec:intro}
The atmospheric compositions of giant planets carry information of their formation and evolution history \citep{Madhusudhan19}. Beyond hydrogen, carbon and oxygen are the two most measurable elements in planetary atmospheres, because they are highly abundant and the C/O ratio has large effects on the atmospheric chemistry of giant planets \citep{Fortney08,Moses13}. In the solar system, the C/H ratio in the atmospheres of Jupiter and Saturn appears to be enhanced by a factor of a few compared to the solar value \citep{Owen99, Atreya05}. Beyond the solar system, super-stellar C/H ratios have been reported in atmospheres of several gas giant exoplanets \citep{Madhusudhan11,Lee13, Lavie17}.
Current studies show that atmospheres of gas giant exoplanets have a wide range of measured C/O ratios \citep{Madhusudhan12,Lee13,Moses13,Line14, Brogi14}; however there is a preponderance of super-stellar C/O ratios, albeit with large uncertainties \citep{Brewer17}. 

These C/H and C/O measurements offer important constraints to test planet formation models \citep{Madhusudhan12, Cridland19}. In the core-accretion formation scenario, the atmospheric composition of a giant planet is initially set by the composition of the gas within the natal disk \citep{pollack96}. For a {\it static} disk, the C/H ratio in the disk gas is expected to be always substellar as 25-75\% of the carbon is in refractory materials \citep{pollack94, Mishra15}. The sublimation/destruction temperature of the carbonaceous refractory grains is at least $> 350~K$ \citep{Gail17}, which lies interior 1\,au in protoplanetary disks \citep{Trilling12}. Therefore, for the majority of the disk the gas phase C/H ratio starts at a substellar ratio and decrementally decreases with distance from the star, as various carbon carriers subsequently freeze-out beyond their snowlines \citep{Oberg11,cridland16, Eistrup16}. Therefore, a super-stellar C/H ratio in planetary atmospheres is usually attributed to contamination of planetesimals or mixing with planetary core materials \citep{Owen99}.

However, contamination or mixing from solids cannot explain super-stellar C/O ratios seen in various exoplanets. This is because solids are more enriched by oxygen than carbon. A possible solution is that the protoplanet accretes gas with an elevated C/H ratio that is enriched by CO ice sublimation from a large amount of icy pebble drifting into the CO snowline \citep{Oberg16,booth17}. In fact, numerical simulations of the formation and drifting of pebbles in disks have long predicted elevated C/H or O/H ratios in the gas inside snowlines \citep[e.g.,][]{Cuzzi04, Ciesla06, Stammler17, Krijt18}. But no previous observation was able to confirm the existence of C/H enriched gas in protoplanetary disks. 

 The HD 163296 system presents a unique target to study the spatial distribution of C/H ratio in a protostar-disk system. The C/H ratio of its stellar photosphere has been measured to be 1.5$^{+1.2}_{-0.7}\times$10$^{-4}$ \citep{Folsom12,Jermyn18}. The total hydrogen mass of the disk is constrained by the upper limit of HD (1-0) line flux \citep{Kama19}. CO is one of the main carriers of carbon in protoplanetary disks and is expected to taken 25-60\% of total stellar carbon budget \citep{Oberg11}. Here we use multiple CO isotopologue $J$=2-1 line spectra to constrain the carbon budget in the disk.

\section{Observations} \label{sec:obs}

 The observations were carried out with the NOEMA interferometer on March 09 and 20, 2019. The total on-source integration time was 3.7 hours. The observations used the wide-band correlator PolyFix that has an instantaneous dual-polarization coverage of 15.5\,GHz bandwidth at a fixed resolution of 2000\,kHz. In addition, higher spectral resolution chunks were set at the line centers of \ce, \cseven, and \ccc~(2-1) with a resolution of 65\,kHz. The baseline lengths were between 24 to 368\,m. Nearby quasars (1730-130 and 1830-210) were observed between science targets to calibrate the complex antenna gains. The absolute flux calibrator was MWC 349. Data calibration and imaging were done using the GILDAS software. We re-binned data to a channel width of 0.5\,\kms~to enhance signal-to-noise ratios. After a uniform weighting, the synthesized beam is 5\farcs 3$\times$1\farcs 5 with a noise level of 8\,mJy beam$^{-1}$. The absolute flux uncertainty is expected to be 15-20\% and the 1.3\,mm continuum flux is consistent with literature values within 15\%.

 All isotopologue CO lines were detected. The integrated line fluxes are listed in Table~\ref{tab:line_flux}. Line fluxes were integrated from greater than 3$\sigma$ region between 0--12\,\kms, except for the weakest \ccc~line. The \ccc~line flux was integrated using the \cseven~line $>$3$\sigma$ region as a mask.

\begin{figure*}[!ht]
\epsscale{1.2}
%\vspace{-0.5cm}
\plotone{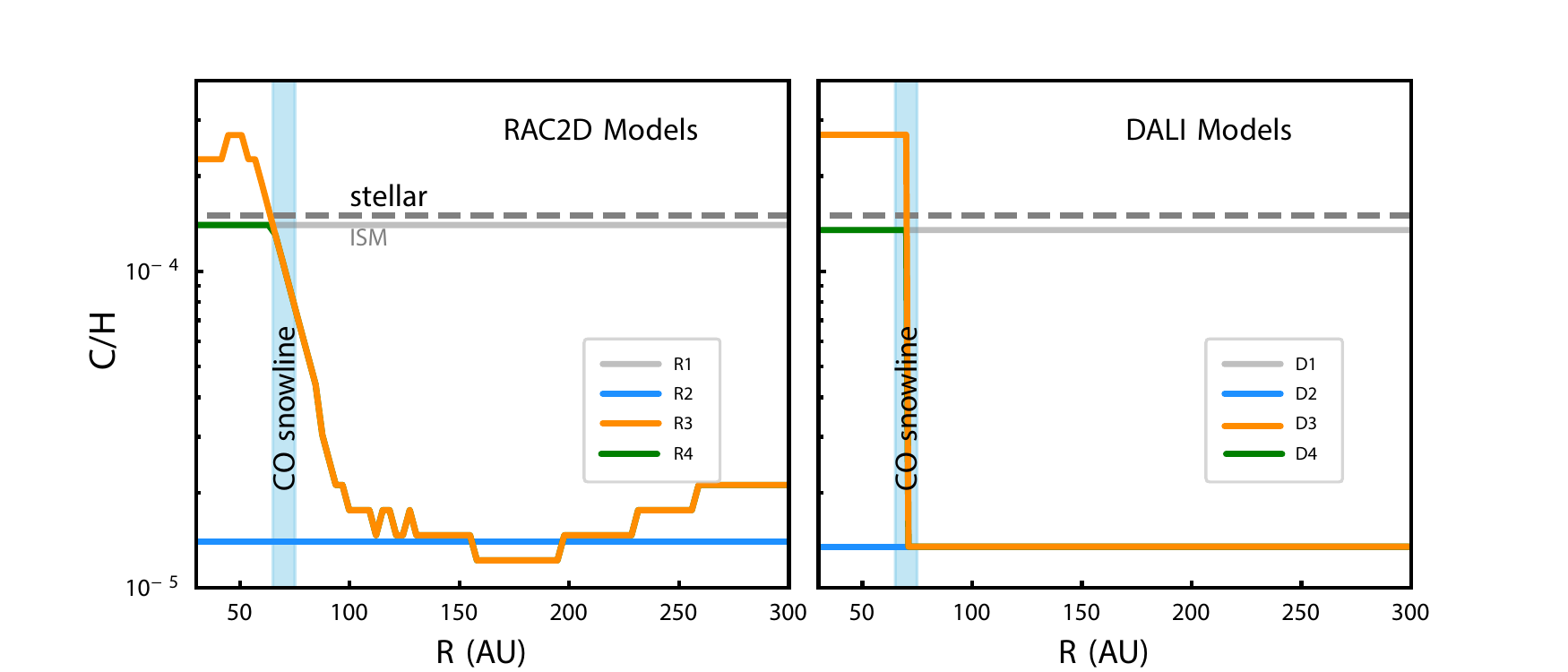}
%\vspace{-1.cm}
\caption{C/H profiles used in the RAC2D and DALI models. %The expected C/H in disk gas is 0.5$\times$stellar ratio, assuming the disk is static and 50\% carbon in refractory materials \citep{pollack94}. 
\label{fig:ch_profiles}}
\vspace{0.6cm}
\end{figure*}

\begin{figure*}[ht]
\epsscale{1.25}
%\vspace{-1.4cm}
\plotone{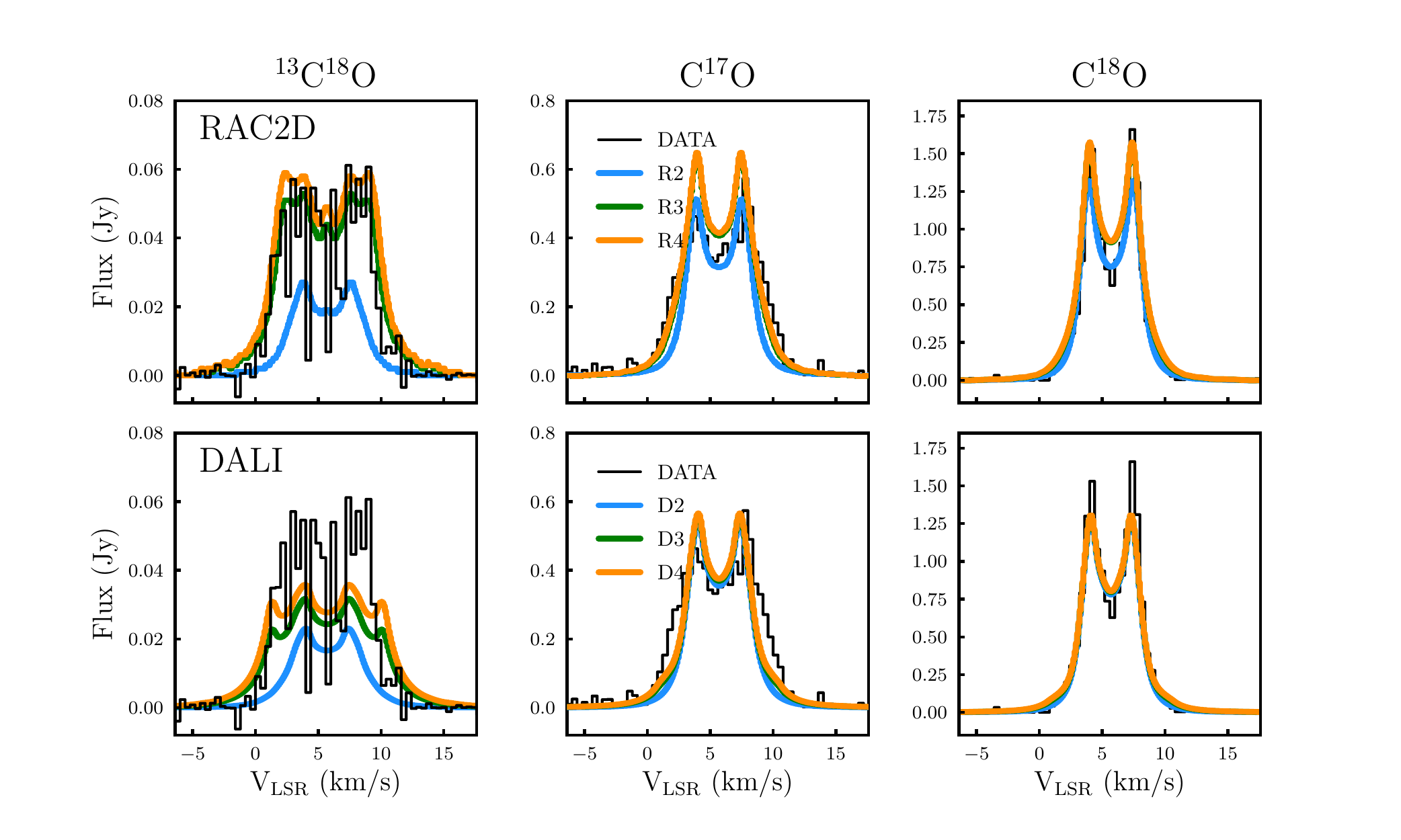}
%\vspace{-1.4cm}
\caption{Comparison of RAC2D and DALI model spectra with observations. The best-fit models are the R3-4 and D3-4 models which require a C/H ratio $\ge$ the stellar ratio inside the CO snowline. \label{fig:model_spec}}
\vspace{0.6cm}
\end{figure*}

\section{Methods} \label{sec:methods}

\begin{deluxetable*}{lccccccccc}
\tablecaption{Line fluxes: data vs. models \label{tab:line_flux}}
\tablehead{\colhead{Line} & \colhead{obs}&
 \multicolumn{4}{c}{RAC2D models} &  \multicolumn{4}{c}{DALI models}\\
 \colhead{} & \colhead{(Jy\,\kms)} & \colhead{R1}&\colhead{R2}&\colhead{R3}&\colhead{R4}&\colhead{D1}&\colhead{D2}&\colhead{D3}&\colhead{R4}
}
\startdata
\ccc\,(2-1)	&	0.37$\pm$0.03	&	1.29	&	0.16	&	0.51 & 0.44	&	1.18	&	0.15	&	0.36 & 0.29	\\
\cseven\,(2-1)	&	3.36$\pm$0.03	&	11.27	&	2.59	&	3.67 & 3.56 	&	12.01	&	2.83	&	3.24 & 3.13	\\
\ce\,(2-1)	&	6.34$\pm$0.04	&	18.64	&	5.98	&	7.50 &7.38	&	19.19	&	5.88	&	6.33 &6.20	\\
\hline
HD\,(1-0) & $\le$67 & \multicolumn{4}{c}{108} & \multicolumn{4}{c}{67}\\
\enddata
\end{deluxetable*}

 We match the observed CO line spectra with thermo-chemical models to constrain the spatial distribution of the C/H ratio in the HD 163296 disk. To make a robust constraint, we employ two thermo-chemical codes, RAC2D and DALI, which were developed independently \citep{du14,Bruderer12,bruderer13}.  Both have been used in molecular line studies of protoplanetary disks \citep[e.g.,][]{bergin16,Bosman18,zhang19}.
 
 \subsection{Common setups of both codes}
 We employ the gas and density structures from \citet{isella16}. The model contains two populations of grains, a population of small grains following the gas, ranging from 0.005-1\,$\mu$m and a second population of large grains, ranging from 0.005-1000\,$\mu$m. The total disk mass is 0.14\,M$_\odot$, assuming a gas-to-dust ratio of 100. 
 
 The chemical structure is computed by starting with an ISM level of elemental abundances across the whole disk\footnote{Please see elemental abundances in the Table\,1 of \citet{du14}.} and then letting the chemistry and gas temperature self-consistently evolve for 1\,Myr. The initial C/H abundance is set to an ISM ratio of 1.4$\times$10$^{-4}$, with all carbon in CO gas. Chemical processes can turn CO into other carbon species and the efficiency is sensitive to the ionization rate in the disk, especially the cosmic-ray rate \citep{Schwarz18, Bosman18}. We use a cosmic-ray rate of 1.36$\times$10$^{-18}$\,s$^{-1}$; this is consistent with a reduced rate through the influence of stellar winds \citep{cleeves13a}. After 1\,Myr, most of the carbon is still in CO and only less than 10\% of carbon has been processed into other carbon species.

\subsection{Thermo-chemical set 1: RAC2D models}
For the first set of models, we start with the RAC2D model of the HD 163296 disk by \citet{zhang19}. This baseline model matches the Spectral Energy Distribution (SED) of the HD 163296 system. It has a mid-plane CO snowline at 70\,au, consistent with earlier constraints on the CO snowline location from spatially resolved images of \ce~and N$_2$H$^+$ lines \citep{Qi15}. 

Previous studies have reported that the CO-to-H abundance in the HD 163296 may be lower than the ISM value of 1.4$\times$10$^{-4}$ \citep{rosenfeld13b,williams14}. Furthermore, \citet{zhang19} showed that the spatially resolved \ce~(2-1) images could not be reproduced by uniformly reducing the CO abundance across the whole disk. Here, we use four sets of models: 
\begin{itemize}
  \item R1: the CO gas abundance structure from the baseline RAC2D model, where the C/H abundance is 1.4$\times$10$^{-4}$ across the whole disk.
  \item R2: we reduce the baseline CO abundance structure by a factor of 10, i.e., a C/H abundance of 1.4$\times$10$^{-5}$ across the whole disk.
  \item R3: we use the radial dependent CO depletion profile derived by \citet{zhang19}, which is required to reproduce spatially resolved \ce~(2-1) line images. The most prominent feature of the profile is the C/H abundance rapidly increases from $\sim$ 0.1$\times$ISM outside the CO snowline to 2$\times$ISM value inside the snowline.
  \item R4: the same as R3, except that the C/H abundance is the ISM value inside the mid-plane CO snowline. 
\end{itemize}
 The detailed profiles are shown in Figure~\ref{fig:ch_profiles}. We then generate model spectra of CO isotopologue lines using the ray-tracing module of the RAC2D code. The isotopologue abundance ratios are set to the local ISM CO abundance ratios of $^{18}$O/$^{17}$O=3.6, $^{12}$C/$^{18}$C = 557, and $^{12}$C/$^{13}$C=69 \citep{wilson99}.

\subsection{Thermo-chemical set 2: DALI models}
% The DALI models \citep{Bruderer12, bruderer13} use the density structure from \citet{isella16}. The model contains two grain populations, small grains following the gas, ranging from 0.005-1 $\mu$m and large grains, ranging from 0.005-1000 $\mu$m \citep[Opacities can be found in ][]{bruderer2015}. ISM elemental abundances are used to calculated the temperature structure and the initial CO abundance structure. 

Similar to RAC2D models, we vary the CO abundance structure from a baseline DALI model to compare with observations. The four sets of models are: (D1) baseline model, with a C/H ratio of 1.35$\times$10$^{-4}$; (D2) all CO abundance depleted by a factor of ten; (D3) at radii larger than the mid-plane CO ice line ($T_{\mathrm{mid}} < 21$ K) the CO abundance is dropped by a factor ten, and in the region within the CO ice line, the CO abundance is enhanced by a factor of two; (D4) the same as D3, except the C/H inside the snowline is an ISM ratio.

\section{Results}
\subsection{Models vs. Observations}
The model line spectra and integrated fluxes are shown in Figure~\ref{fig:model_spec} and Table~\ref{tab:line_flux}. RAC2D and DALI models show consistent results, typically with less than 15\% differences in line fluxes. 

R1 and D1 models have a uniform ISM level of C/H ratio across the whole disk. These models overproduce line fluxes by a factor of $\sim$3 compared with observations. R2 and D2 models are uniformly depleted disk models (0.1$\times$ ISM C/H ratio), these models match the integrated fluxes of \ce~and \cseven~(2-1) observations. The model \ce~ line profiles also match well with observations, but \cseven~models are slightly narrower than the observations. This is most likely the result of hyperfine line splitting in the  \cseven~(2-1) transition, which is not fully implemented in our ray-tracing code \citep{Mangum15}. The R2 and D2 models underpredict the \ccc~(2-1) line flux by a factor of 2.4, and the model line profiles have a FWHM of only $\sim$6\,\kms, much narrower than the observed FWHM of $\sim$10\,\kms.  This strongly hints that the \ccc{} line emission originates from velocities consistent with the inner $\sim 70$~au; i.e. inside the CO snowline. 

The R3-4 and D3-4 models represent distributions with an 1-2$\times$ ISM level of C/H ratio inside the CO snowline and a depleted C/H ratio gas of outer region. These models provide the best over-all match to observations, and only these can match the observed \ccc~(2-1) line flux and the general profile. We note that these models over produce the line intensities at high velocity channels that are $\ge$5\,\kms~offset from the stellar velocity. One possible solution is that the C/H ratio inside $\sim$ 40\,au is lower than the ISM level. But the existing observations do not have sufficient signal-to-noise ratio or spatial resolution to constrain the detailed C/H profile inside 40\,au. 

\subsection{Result Robustness}
The HD 163296 disk is known to have substructures (e.g., gaps/rings) in its 1.3\,mm continuum emission \citep{zhang16,isella16}. To test the effects of substructures in the continuum, we run additional models that include dust substructures and find these do not change our conclusion.

All of our constraints on the C/H ratio in disk depend on the total disk mass. In Table~\ref{tab:line_flux}, we show that our model HD (1-0) line fluxes are consistent or slightly overproduce the upper limit reported by \citet{Kama19}. This suggests that the total disk gas mass (0.14\,M$_{\odot}$) in our models is an upper limit. Therefore our estimation of the excess of C/H inside the CO snowline is a  robust lower limit and the actual C/H can be even higher. 

We note that there is a detection of $^{13}$C$^{17}$O (3-2) line in the HD 163296 disk with the match-filter method \citep{Booth19}. Although the line flux is uncertain, the detection suggests a high column of CO gas inside the CO snowline, which is consistent with our results. 

In short, we find that in the HD 163296 protoplanetary disk: (1) the gas between 40-70\,au has a C/H ratio of 1-2$\times$the ISM value; (2) the gas outside 70\,au has a C/H ratio of 0.1$\times$ISM value.

\section{Discussion} \label{sec:discussion}
The stellar carbon abundance represents the total amount of carbon available in the bulk materials of protoplanetary disks. Here we compare our results to the carbon abundance of the HD 163296 star. Its stellar C/H ratio is measured as 1.5$^{+1.2}_{-0.7}\times$10$^{-4}$ \citep{Folsom12,Jermyn18} which is about half of that in the Sun \citep [2.7$^{+0.3}_{-0.3}\times10^{-4}$,][]{Asplund09}. Therefore HD 163296 system started with relatively carbon poor materials than the ISM.

\subsection{Fraction of carbon in CO}
\label{subsec:co_budget}
Our C/H constraint is essentially a CO/H ratio, as our chemical models have $\ge$90\% of carbon in CO and the \ccc~(2-1) line spectrum does not sensitive to the region inside the CO$_2$ snowline ($<$40\,au). Therefore we discuss to what fraction CO can take up the total carbon budget. For reference points, we use carbon budgets measured in ISM, molecular clouds, protostellar cores, and solar system objects. 

In the ISM, ~50\% of the cosmic carbon is locked in refractory grain materials \citep{draine03}. More recently, \citet{Mishra15} studied the abundance of carbon grains using the UV extinction along 16 Galactic sightlines. They found refractory carbon grains, on average, take 50\% of the total carbon budget, varying between 25-75\% in their sample. These refractory materials are expected to survive from ISM to protoplanetary disks, because abundant  carbon-rich grains are found in-situ collections of dust in coma of comet Halley and 67P/Churyumov \citep{Jessberger88, Bardyn17}.  Theoretical studies also suggest that the processing of carbon grains is inefficient in protoplanetary disks \citep{anderson17,Klarmann18}. Further, our own thermal models suggest that the refractory carbon grain destruction zone will lie interior to 1\,au. We note that the bulk carbon abundance in chondrites is significantly lower than that of comets \citep[e.g.,][]{bergin15}; this  likely  requires  refractory carbon grain destruction in the inner solar system that would release carbon to the gas. But chondrites likely formed inside a few au and thus their compositions do not represent that of the radial region of $>$20\,au we study here.

For volatile carbon-materials,  measurements of comets and protostellar cores provide the best constraints. \citet{Pontoppidan06} studied ice absorption features towards 5 young stellar sources in the Oph-F core and reported CO-to-CO$_2$ abundance ratios varying between 0.48 and 3. In a much larger sample, \citet{oberg11_ice} studied 63 young stellar objects. They found the combination of CO and CO$_2$ dominate the carbon budgets in ice species --- on average, these two take 87\% of the total carbon in ice species in low-mass stellar objects. The medium CO-to-CO$_2$ ratio is unity, and for most of the low-mass sources, the ratio varies between 0.4 and 2.5. Measurements of comets show similar CO-to-CO$_2$ ratios \citep{mumma11}.  All these studies suggest that CO only takes a fraction of the total carbon budget in protoplanetary disks. 

To evaluate the significance of the CO excess in the HD 163296 disk, we compare our constraint of CO abundance with two cases of CO fraction in the carbon budget. The first case is based on the average composition of the major carbon species discussed above. In this case, refractory carbon grains, CO$_2$, and CO take 50\%, 25\%, and 25\% of the total stellar carbon budget, respectively. We note that including other minor carbon species, such as CH$_4$ and CH$_3$OH, will make the CO fraction even lower. For the second case, we consider a case that maximizes the carbon fraction in CO. We adopt the lowest carbon grain fraction of 25\% found in diffuse ISM \citep{Mishra15}, and a high CO-to-CO$_2$ abundance ratio of 3 \citep{Pontoppidan06,oberg11_ice}. Here refractory carbon grains, CO$_2$, and CO take 25\%, 18.75\%, and 56.25\% of the total stellar carbon budget. 

\begin{figure}[!t]
\epsscale{1.2}
\vspace{0.2cm}
\plotone{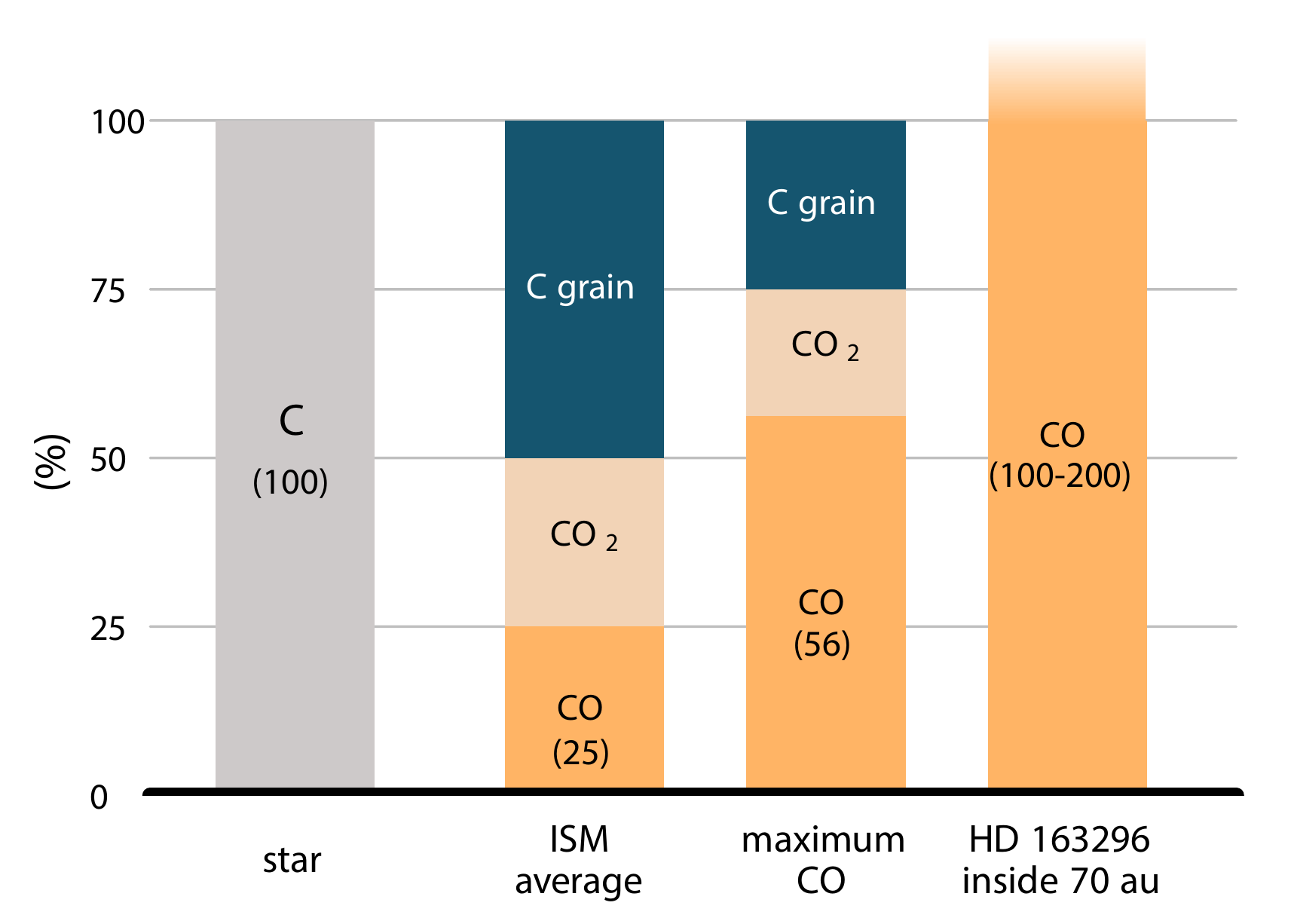}
%\vspace{0.3cm}
\caption{Comparison between the expected CO abundance with the observed CO abundance inside 70\,au of the HD 163296 disk. The ISM average case represents average ratios of refractory carbon grains, CO$_2$, and CO seen in ISM, protostellar cores, and comets. The maximum CO case shows the highest fraction of carbon in CO, by adopting lower limits of other major carbon species. \label{fig:c_budget}}
\vspace{0.6cm}
\end{figure}

In Figure~\ref{fig:c_budget}, we compare the CO fractions in these two cases with the CO abundance requirement of the inner 70 au of the HD 163296 disk. Compared with the best estimation of the stellar carbon abundance of 1.5$\times$10$^{-4}$, our constraint of the CO abundance inside 70 au  exceeds the expected value e by a factor of 4-8 in the case of the ISM average composition. In the case of the maximal CO fraction, the observational CO abundance is a factor of 1.8-3.6 higher than the expected abundance. However, we note that in the maximal CO fraction case, the lower limit of observational CO abundance inside 70\,au is still consistent with upper limit of the stellar carbon abundance.  In summary, although we cannot completely rule out the possibility of a normal C/H ratio in the gas inside 70\,au, our results suggest that the gas inside 70\,au region of HD 163296 disk likely has an elevated C/H ratio that exceeds the expected value by a factor of 1.8-8.

\subsection{ Excess C/H as a test of pebble drift models}
\label{subsec:pebble_drifting}

Our results suggest the gas inside 70\,au of the HD 163296 disk has an elevated C/H ratio. Here we discuss the implication of this result in the context of the pebble accretion framework. 

The mass flux of pebble drifting from the outer to the inner disk is a crucial parameter of the pebble accretion models. Here we do an order of magnitude estimation on the pebble mass flux in the HD 163296 disk. 
To increase the CO/H$_2$ ratio in the gas inside 70\,au by a factor of 1.8 to 8, it requires to increase the solid mass initially inside the 70\,au region by the same factor, with a supply of CO-ice coated grains from the outer disk. In our HD 163296 model, this requires that 150-600\,M$_\oplus$ of pebbles drift into the inner 70\,au region within the disk lifetime. Given the age of the disk is estimated to be 5-10\,Myr, this leads to a pebble mass flux of $\sim$15-60\,M$_\oplus$/Myr. This mass flux is comparable to the $\sim$95\,M$_\oplus$/Myr derived from the analytical pebble drifting model of \citet{Lambrechts14}. 

An elevated C/H gas ratio inside the CO snowline has long been predicted by models that include dust drift \citep[][]{Cuzzi04}. More comprehensive simulations consider icy pebble formation, settling, and drifting with CO sublimation in a global disk setup \citep{booth17,Stammler17,Krijt18}. In general, these models predict the C/H gas ratio inside the CO snowline can be elevated to 1--10 times of the initial ratio, while the detailed radial distribution depends on various parameters, including viscosity, diffusion rate, and disk sizes. Although our data cannot constrain the detailed radial profile of the C/H ratio inside the CO snowline, we find that the gas inside $\sim$40\,au might have a lower C/H ratio than the gas between 40-70\,au, which is consistent with model predictions that the elevated C/H ratio is most prominent in region just inside the CO snowline.

\begin{figure}[ht]
\epsscale{1.15}
\vspace{0.2cm}
\plotone{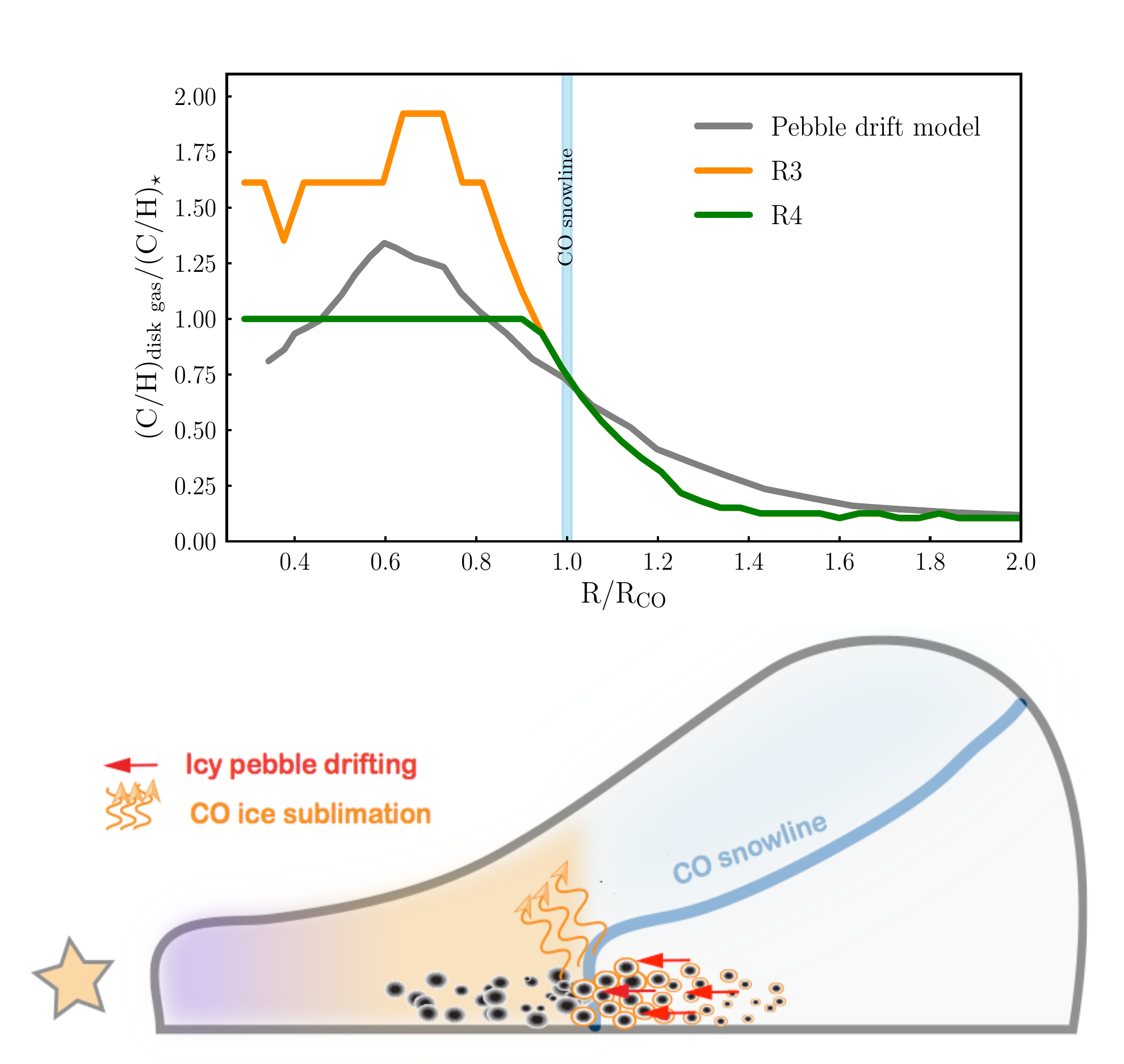}
\vspace{0.2cm}
\caption{Top: comparison of our best-fit models with pebble drift simulations from \citet{Krijt18}. To rescale the CO/H$_2$ ratio from \citet{Krijt18}, we take the maximum abundance that 56\% of carbon in CO (see the discussion in section \ref{subsec:co_budget}).  Bottom: an illustrative figure that the disk gas in the CO snowline is enriched in C/H by ice pebble drift.   \label{fig:comparison}}
\vspace{0.6cm}
\end{figure}

To make a more quantitative comparison, in Figure~\ref{fig:comparison} we compare our best-fit models with predictions of \citet{Krijt18} . \citet{Krijt18} is the only 2-dimensional simulation that included the depletion of CO gas in the warm molecular layer outside the mid-plane CO snowline, which gives the closest theoretical comparison to our constraints on the C/H ratio in gas phase. Even the \citet{Krijt18} models were for a generic disk, our best-fit models match the general profile within a factor of two. 

However, we do not have sufficient spatial resolution or signal-to-noise ratio to constrain the detailed radial profile of the C/H ratio inside the CO snowline. Higher spatial resolution ALMA images would be necessary to accurately constrain the elevated C/H ratio and quantitatively test current pebble drift simulations.

\subsection{ C/H and C/O ratios as an indicator of planet formation location and history}
\label{subsec:giant_planets}

The chemical composition of a giant planet depends on the relative amounts of disk gas and solids accreted during its formation, the compositions of which in turn depend on time and location. On top of that, planetesimal delivery and core-envelope mixing can further alter the atmospheric composition. 

Given the complexity of the formation processes, it is challenging to uniquely attribute an elemental ratio in a planetary atmosphere to a single process. For example, a super-stellar C/H atmosphere can be produced by accreting C/H enrich disk gas, planetesimal contamination, or core erosion. Similarly, a C/O ratio of unity can be from accreting gas between the CO$_2$ and CO snowlines, or from the CO enriched gas due to pebble drift into the CO snowline. The combination of several elemental ratios can be more constrictive. \citet{Oberg16} and \citet{booth17} propose that if a giant planet atmosphere has both a super-stellar C/H and a C/O ratio $\sim$1, it is a unique indicator of the planetary atmosphere formed from gas enriched by ice pebbles drift into the CO snowline. 

It is also of particular interest to look at the case of the HD 163296 disk, as its abundant substructures suggest ongoing planet formation in the disk. Recent gas kinematic studies revealed signatures of gas flows into locations of three gaps opened by three giant planets in the HD 163296 disk \citep{Teague18a,Teague19Nat}. All three accreting planets are located outside the CO snowline, where we find the C/H ratio in gas phase is only 0.1$\times$ stellar value. If these planets are still accreting a significant amount of their atmospheres, the high depletion C/H gas may leave a low C/H ratio in their final planetary atmospheres. Inside the CO snowline, the disk shows two additional gaps at 10 and 48\,au in its 1.3\,mm continuum image, which are characterized as gaps opened by Jupiter-mass planets \citep{Zhang_S18}. Existing CO gas observations do not have sufficient resolution to show if gas also flows into these inner gaps. But if these planets are also accreting, the C/H ratio in their atmospheres would be an order of magnitude higher than these planets accreting atmospheres from the gas outside the CO snowline. Although highly speculative, the HD 163296 disk hints a possibility that giant planets form inside or outside of the CO snowline may carry distinctive C/H ratios in their atmospheres as a birthmark.

\section{Summary}\label{sec:summary}
We report the first detection of excess C/H in the gas of a protoplanetary disk. In the HD 163296 disk, we find its gas just inside the CO snowline ($\sim$70\,au) has a C/H ratio of 1.4-2.8$\times$10$^{-4}$, which is  1-2$\times$ of the stellar C/H ratio of 1.5$^{+1.2}_{-0.7}\times$10$^{-4}$.   This gas C/H inside 70\,au is significantly higher than the expected ratio, as only 25-60\%~of the stellar carbon budget should be in gas at these radii. Although existing observations cannot completely rule out the case of a normal C/H ratio inside 70 au, the most probable solution is an elevated C/H ratio of 1.8-8 times higher than the expected ratio. The C/H enriched gas is consistent with predictions of a large amount of icy pebble drift across the CO snowline. 
\newpage

\acknowledgments
This work is based on observations carried out under project number W18AB with the IRAM NOEMA Interferometer. IRAM is supported by INSU/CNRS (France), MPG (Germany) and IGN (Spain). We thank the IRAM staff member Jan-Martin Winters for assistance of observations and data calibrations.
K.Z. acknowledges the support of NASA through Hubble Fellowship grant HST-HF2-51401.001 awarded by the Space Telescope Science Institute, which is operated by the Association of Universities for Research in Astronomy, Inc., for NASA, under contract NAS5-26555.   EAB acknowledges support from NSF Grant\#1907653.

%% To help institutions obtain information on the effectiveness of their 
%% telescopes the AAS Journals has created a group of keywords for telescope 
%% facilities.
%
%% Following the acknowledgments section, use the following syntax and the
%% \facility{} or \facilities{} macros to list the keywords of facilities used 
%% in the research for the paper.  Each keyword is check against the master 
%% list during copy editing.  Individual instruments can be provided in 
%% parentheses, after the keyword, but they are not verified.

%\vspace{35mm}
\newpage
\facilities{NOEMA(PolyFix)}

%% Similar to \facility{}, there is the optional \software command to allow 
%% authors a place to specify which programs were used during the creation of 
%% the manuscript. Authors should list each code and include either a
%% citation or url to the code inside ()s when available.

\software{RAC2D \citep{du14}, DALI \citep{Bruderer12,bruderer13},
          GILDAS \footnote{See http://www.iram.fr/IRAMFR/GILDAS for more information about the GILDAS softwares.}, astropy \citep{Astropy13}}

\bibliographystyle{aasjournal}
\bibliography{lib}{}

%% This command is needed to show the entire author+affiliation list when
%% the collaboration and author truncation commands are used.  It has to
%% go at the end of the manuscript.
%\allauthors

%% Include this line if you are using the \added, \replaced, \deleted
%% commands to see a summary list of all changes at the end of the article.
%\listofchanges

\end{document}